\documentclass[journal,10pt]{IEEEtran}
\IEEEoverridecommandlockouts
\usepackage{cite}
\usepackage{indentfirst}
\usepackage{multirow}
\usepackage{booktabs}
\usepackage{makecell}
\usepackage{float}
\usepackage{amsmath,amssymb,amsfonts}
\usepackage{algorithmic}
\usepackage{graphicx}
\usepackage{graphicx,subfigure}
\graphicspath{{./figures/}}
\usepackage{textcomp}
\usepackage{xcolor}
\usepackage{subfigure}
\usepackage{booktabs}
\usepackage{bm}
\usepackage{amsthm}
\usepackage{epstopdf}
\usepackage{diagbox} 
\usepackage{booktabs} 
\usepackage[lined,boxed,commentsnumbered,ruled,linesnumbered]{algorithm2e}
\def\BibTeX{{\rm B\kern-.05em{\sc i\kern-.025em b}\kern-.08em
		T\kern-.1667em\lower.7ex\hbox{E}\kern-.125emX}}

\begin{document}
	\title{ODE-Former for Mobile Channel Prediction:\\ A Novel Learning Structure Leveraging The Physics Continuity}
	\author{Zhuoran~Xiao,~\IEEEmembership{Member,~IEEE, }
		\thanks{Zhuoran Xiao (e-mail: zhuoran.xiao@nokia-sbell.com) is with Nokia Bell Labs, Shanghai 201206, China.}

%
	}

	\maketitle

	\begin{abstract}
		Obtaining accurate channel state information (CSI) is crucial and challenging for multiple-input multiple-output (MIMO) wireless communication systems. With the increasing antenna scale and user mobility, traditional channel estimation approaches suffer greatly from high signaling overhead and channel aging problems. By exploring the intrinsic correlation among a set of historical CSI instances, channel prediction is proven to increase the CSI accuracy while lowering the signaling overhead significantly. Existing works view this problem as a regular discrete sequence prediction task while ignoring the unique physics property of wireless channels. This letter proposes a novel former-like learning structure based on neural ordinary differential equations (NODEs) inclusively designed for accurate and flexible channel prediction.  The proposed network aims to represent wireless channels' implicit physics spatial-temporal continuity by integrating the Neural ODE into a former-like learning structure. Our proposed method impeccably fits channel matrices' mathematics features and enjoys solid network interpretability. Experimental results show that the proposed learning approach outperforms existing methods from the perspective of accuracy, flexibility, and robustness.
	\end{abstract}


	\section{Introduction} \label{intro}

Accurately obtaining Channel State Information is indispensable for effectively functioning wireless communication systems. In existing systems, CSI is initially obtained at the receiver with the help of pre-defined pilots by channel estimation algorithm and then feedback to the BS \cite{seq2seq_feedback}. However, at least three potential drawbacks exist to this approach for the mobile scenario. On the one hand, with the fast variability of mobile channels, channel estimation should be conducted more frequently to ensure the accuracy of the estimated channel. Meanwhile, with the fast-growing antenna scale, the number of parameters to be estimated greatly increases, which undoubtedly adds to the signaling overhead \cite{8395053}. Moreover, due to the nonnegligible transmission delay for the BS to receive feedback data, there exist mismatches between the feedback channel and the instantaneous channel, which will inevitably cause the loss of accuracy \cite{9791407}.

By leveraging the temporal and spatial correlation between the instantaneous and historical channels, channel prediction is an effective approach to obtaining more accurate CSI against heavy signaling overhead and processing delay. Conventional prediction approaches mainly include the parametric model and auto-regressive method \cite{2000Long}. Due to the gap between the prediction model and the real wireless channel, the statistical modeling-based prediction is generally inaccurate in practical systems. By applying machine learning-based algorithms such as recurrent neural network (RNN) \cite{2002Recurrent}, long short-term memory (LSTM) \cite{2020Recurrent} and transformer \cite{9832933}, recent works achieve better performance. However, these works simply regard the wireless channel prediction task as a regular sequence prediction task while ignoring the unique intrinsic properties of the wireless channel, which leads to insufficiency. Besides, these methods require a strict equal interval of the obtained CSI sequence, which greatly constrains the flexibility of practical usage.

The variation process of a mobile channel is a continuous physics process with absolute causal consistency instead of a discrete generation process. The continuity mainly comes from three reasons: the propagation of electromagnetic waves is a continuous physics process governed by Maxwell's equations, the UE's moving process is continuous, and the scattering surface is continuous. Consequently, it is a more advantageous approach to model the evolution of the mobile channel over time rather than attempting to synthesize it directly. Authors in \cite{10351043} propose using the Neural ODE to represent the implicit channel variation process between each RNN cell. However, due to the inherent limitations of RNN, such as long-term forgetting and difficulty scaling the parameter size of the network, it still needs to be more accurate for practical usage.

To solve the problems mentioned above, we introduce the inclusively designed ODE-Former network. This innovative network structure tries to incorporate physical channels' characteristics into the network's structural design. There are several key highlights of the network design. Firstly, three layers are designed to jointly represent the implicit physics process of component path response variation. Secondly, motivated by the stacking structure of the Transformer, we achieve to stack our proposed ODE-former block while not losing convergence, which significantly enhances the network learning capability compared to the original Neural ODE network. Thirdly, our proposed network is naturally suitable for parallel computation compared to existing works such as ODE-RNN, which require sequential computation for few-slots prediction. Thus, channels of the future time slots can be calculated simultaneously, significantly reducing inference latency.

The remainder of this letter is organized as follows. The system model and problem formulation are in section \ref{system}. The key motivation and network design are given in section \ref{net}. Section \ref{scene} introduces our experiment scene setup. Numerical results that evaluate our proposed approach against state-of-the-art (SOTA) from diverse perspectives are presented in section \ref{performance}. Section \ref{conclusion} draws our main conclusions.

\section{Problem Formulation} \label{system}
The goal of channel prediction is to forecast the CSI at the current and following time slots by leveraging the CSI information obtained in previous periods. In existing works \cite{2020Recurrent,8904286}, the CSI sequence requires to be uniformly sampled.
Assume that the BS stores estimated CSI in the past $n$ time slots, which can be denoted by $\{ {\bf{H}}[1],{\bf{H}}[2],...,{\bf{H}}[n]\} $. The CSI in the next time slot should be predicted, denoted by $\widehat{\bf{H}}[n + 1]$.
Thus, the CSI prediction problem can be presented as
	\begin{equation}
		\{ {\bf{H}}[1],{\bf{H}}[2],...,{\bf{H}}[n]\}  \to \widehat{\bf{H}}[n + 1].
	\end{equation}

As one advantage of our proposed method, the CSI sequence is no longer required to be evenly sampled. Thus, the process can be written as
	\begin{equation}
		\{ {\bf{H}}[{t_1}],{\bf{H}}[{t_2}],...,{\bf{H}}[{t_n}]\}  \to \widehat{\bf{H}}[{t_{n + 1}}],
	\end{equation}
where $t_x$ is a certain time point and the time sequence $t_1, t_2, \cdots, t_n, t_{n+1}$ can be chosen as any arbitrary sequence.

\section{ODE-Former for Mobile Channel Prediction: Motivation and Learning Structure} \label{net}

\subsection{ODE Representation of Mobile Channel}
The essential fact distinguishing the channel prediction task from other temporal prediction problems is that the channel variation over time is a continuous physics process that can be uniformly represented within a specific scattering environment. As has been proved in \cite{10351043,9978073}, the derivative of the channel matrix with respect to time is only related to the current CSI and the moving direction of UE. Since the prediction interval is of the same order as the channel coherent time, which usually measures in milliseconds, the user can be considered to be undergoing uniform rectilinear motion. Thus, the physics process of CSI variation can be represented with an ODE function as
	\begin{equation}
		{{\partial {\bf{H}}[t]} \over {\partial t}} = f({\bf{{H}}}[t],t).
	\end{equation}
Since the function $f(\cdot)$ is not explicitly known, a neural network can be used to fit it
	\begin{equation}
		f({{\bf{H}}[t]},t) = \Phi({\bf{H}}[t],t,{\boldsymbol\theta} ),
	\end{equation}
where $\Phi(\cdot)$ denotes the network structure and $\boldsymbol\theta$ denotes the parameters. Thus, the pending prediction CSI can be obtained through an integration process, which can be written as
	\begin{equation}
		{\bf{{H}}}[t+\Delta t] = {\bf{{H}}}[t] + \int_{t}^{t+\Delta t}\Phi({\bf{H}}[t'],t',{\boldsymbol\theta}) \,\mathrm{d}t'.
	\end{equation}
In the practical implementation, the integration process can be solved by numerical methods which is called ODE Solver, which can be written as
	\begin{equation}
		{\bf{{H}}}[t+\Delta t] = \text{ODE Solver}(\Phi_{\boldsymbol\theta},{\bf{H}}[t], t, \Delta t).
	\end{equation}

Further, since each scattering path component is independent. We can easily have
	\begin{equation}
		{{\partial {\bf{H}}_p[t]} \over {\partial t}} = f({\bf{H}}_p[t],t),
	\end{equation}
where ${\bf{H}}[t]=\sum\limits_{p = 1}^K {{\bf{H}}_p[t]}$, and $K$ is the total number of propagation paths. Thus, we have
	\begin{equation} \label{eq11}
		{\bf{{H}}}[t+\Delta t] = \sum\limits_{p = 1}^K \left \{ {\bf{H}}_p[t] + \int_{t}^{t+\Delta t}\Phi_p({\bf{H}}_p[t'],t',{\boldsymbol\theta}) \,\mathrm{d}t' \right \} .
	\end{equation}

Due to the high dynamic properties of channel changing, it is difficult and inaccurate to predict the channel at a future time point directly in a discrete way. However, by representing the quasi-static physics changing rule with a neural network, the channel at a future time point can be obtained through the corresponding evolving calculation process in the form of integration, which greatly ensures accuracy.

	\begin{figure*}
		\centering
		\includegraphics[width=0.8\textwidth]{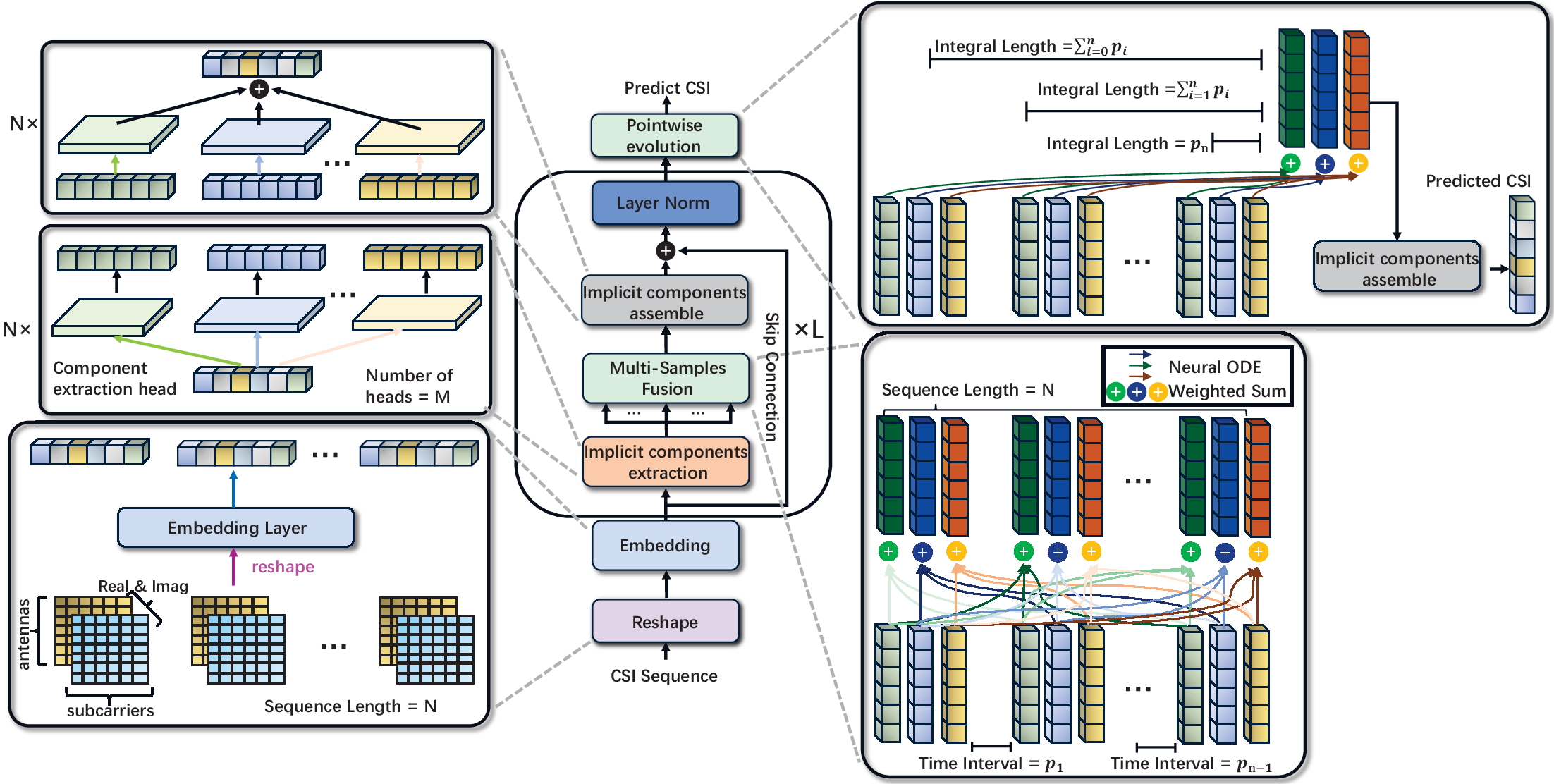}
		\caption{The learning structure of the proposed ODE-Former.}
		\label{network_structure}
		\vspace{-1em}
	\end{figure*}

\subsection{Learning Structure of ODE-Former}
For an intuitive introduction to the network structure shown in Fig. \ref{network_structure}, we describe the functionality of its layers in forward order. We first reshape and scale each original channel matrix in the sequence to a vector. Then, the vectors are fed into an embedding layer, which converts all the vectors to a fixed embedding length $L_{emb}$. All the embedding vectors are then fed into the ODE-Former blocks. Each ODE-Former block comprises four sub-blocks: implicit components extraction block, multi-samples fusion block, implicit components assemble block, and layer norm block. The block designs are motivated by equation (\ref{eq11}).

\subsubsection{Implicit Components Extraction Block}
Firstly, each implicit components extraction block is composed of $M$ heads of MLPs (multi-layer perceptions) and tries to separate independent path response components. It is worth emphasizing that this separation does not need to be mathematically restricted since the $M$ is not always equal to the real number of propagation paths. Basically, more heads could help enhance the network's representation abilities.

\subsubsection{Multi-Samples Fusion Block}
The multi-samples fusion block is the most essential part of this network. For each independent implicit path component vector, we set a neural ODE to represent the temporal evolution process. It is worth mentioning that the integrating calculation of neural ODE is bidirectional. Thus, the earlier channel can also be calculated from the later one. Since we already know the time interval between different sampling points, according to (\ref{eq11}), the channel of a certain time point can be deduced by any other samplings through the ODE solver. So, we can easily fuse the observations of all the time points by the weighted sum of all the deduced results.

For example, suppose the $i$th component the $j$th input sampling points is denoted as $\mathbf{e}_{i,j}$, where $i \in \left \{ x\in \mathbb{Z}|1\le x \le M \right \} $ and $j \in \left \{ x\in \mathbb{Z}|1\le x \le N \right \} $. Then, the corresponding output of this block, denoted as $\mathbf{out}_{i,j}$, can be calculated by

	\begin{equation}
		\mathbf{out}_{i,j} = \sum\limits_{p = 1}^N w_p \left \{ \mathbf{e}_{i,p} + Solver[t_p,t_i,\mathbf{e}_{i,p},\Psi({\boldsymbol\theta}_i)] \right \},
	\end{equation}
where $w_p$ is the weighting coefficient, $\Psi({\boldsymbol\theta}_i)$ is the neural ODE network with parameter ${\boldsymbol\theta}_i$, $t_p$ and $t_i$ denote the sampling time point. There are numerous potential approaches to deciding the value of $w_p$, and here we present two feasible heuristic methods. One is that we can simply set $w_p=\frac{1}{N} $ considering that each sampling point may have equal importance. Alternatively, considering that when using a solver, longer integration leads to an increased accumulation of errors. Thus, the weight of those values with longer integration lengths should be given lower weights. So, we have
	\begin{equation}
		w_p=\frac{e^{1/\left | t_i-t_p \right | } }{\sum\limits_{j = 1}^N e^{1/\left | t_i-t_j \right | }}.
	\end{equation}

\subsubsection{Implicit Components Assemble Block}
The implicit components assemble block adopts a symmetric architecture to that of the implicit components extraction block. This block works to assemble all the path response components. A norm layer is followed behind it to ensure the stability of gradients. Thus, the output dimension of this block is $L_{emb} \times N$. Another critical necessity of this block is that since it reshapes the output of the ODE-Former block back to the embedding size, the skip connection can be achieved, which enables the network to significantly increase its depth without concerns about the vanishing gradient problem.

\subsubsection{Pointwise Evolution Block}
The last block of the network is a special ODE-Former block called a pointwise evolution block. In this block, all the components of the sampling point independently evolve to the final under-predicted channel through integration with different lengths. Then, through weighted sum and an assembled network, the final channel can be obtained. Since the integration process is continuous, we can easily obtain the CSI matrix of any certain time points.

\subsection{Advantages Compared to Existing Works}
The original Neural ODE's integration calculation process is determined with the given initial value. Consequently, the prior CSI sequences cannot be fully leveraged to facilitate accurate prediction. ODE-RNN is a potential solution to this problem. However, there are still three main drawbacks. Firstly, inserting RNN cells in the integration process partly ruins the physics continuity. Secondly, due to the inherent characteristics of the network structure, the RNN cell's learning ability is highly limited, which constrains the achievement of higher accuracy. Thirdly, the structure of RNN networks inherently assigns more influence to samples closer in time to the prediction. However, this characteristic does not align well with the practical realities of channel prediction problems.

Adopting a former-like network structure can significantly enhance the network's learning capacity through block stacking and skip-connection without sacrificing the advantages of the original Neural ODE in representing continuous physical processes. Besides, it solves the problem of long-term forgetting through sampling fusion and is particularly well-suited for parallel computation to reduce computational latency.

Two crucial factors significantly constrain the original Transformer network's capability to model continuous physics processes. The first one is positioning embedding, which is necessary for the Transformer network to catch temporal relationships between inputs. However, since the positioning embedding vector is directly added to the feature vector, the intrinsic characteristic of the internal is significantly broken. The other one is that self-attention calculation is heuristic and does not match the practical data model well. In contrast, in the proposed ODE-Former network, the integration length directly decides the temporal relationship. Therefore, no positioning embedding is needed, and the intrinsic characteristics of the data can be completely preserved. Besides, the proposed network is inclusively designed for the specific problem, which enjoys high interpretability and robustness.

\section{Performance Evaluation}
\subsection{Experiment Settings}
\subsubsection{Scene Setup and Datasets Generation} \label{scene}

	\begin{figure}
		\centering
		\includegraphics[width=0.3\textwidth]{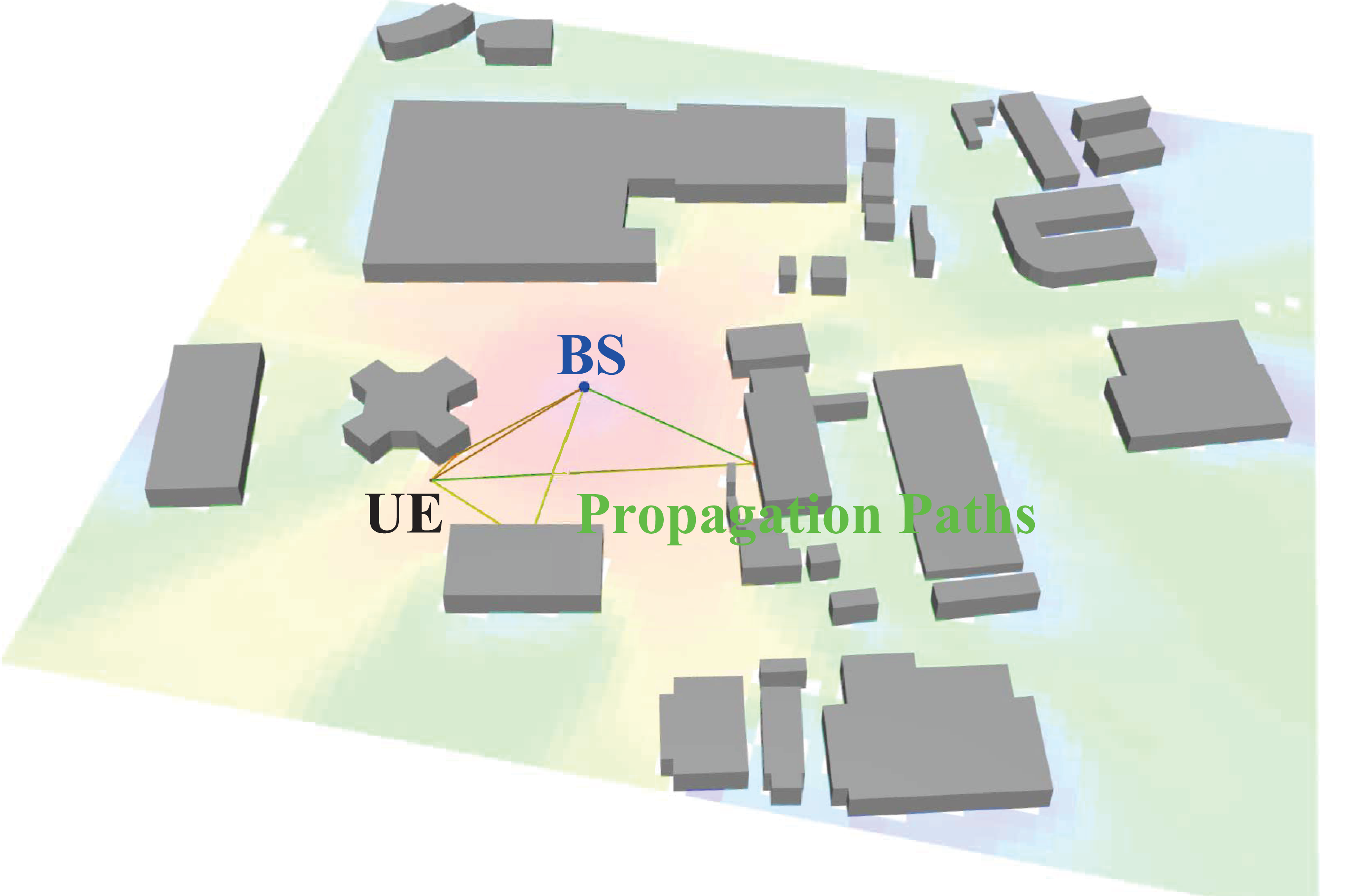}
		\caption{The 3D model of the ray-tracing scene.}
		\label{3d_model}
		\vspace{-1em}
	\end{figure}

To comprehensively evaluate our proposed approach, we have selected two experimental scenarios. We establish a 3D model that accurately represents the scattering environment of the Nokia Shanghai Bell campus. Then, \textit{Winprop} software is used for the ray-tracing calculation. As shown in Fig. \ref{3d_model}, users are sampled in a $700$m × $700$m area where all the building material is cement. The BS with a height of $20$m is equipped with a ULA with $32$ antennas. The central frequency is set to $3.5$GHz. The OFDM bandwidth is $100$MHz, and the number of sub-carriers is $32$. The maximum number of stored paths is $25$. As a supplement, we also employ the WAIC open-source channel prediction dataset, which is derived from real-world sampling in the scenario and contains sequence CSI sampled with different UE speeds. Compared to ray-tracing datasets, this dataset exhibits greater non-ideality, making it more suitable for reflecting the feasibility of the proposed method in practical scenarios. For each experimental setup, we sample $50000$ data sequences for training and $5000$ for testing. The detailed settings are summarized in Table \ref{experiment_setting}.

\begin{table}[h!]
	\caption{Experiment Settings}
	\begin{center}
		\begin{tabular}{ p{2.5cm} p{2.6cm} p{2.3cm} }
			\toprule
			\textbf{Parameters} & \textbf{Ray-tracing scenario} & \textbf{Measured dataset} \\
			\toprule
			Central frequency & $3.5$Ghz & $2$Ghz \\
			Carrier bandwidth & $100$Mhz & $10$Mhz \\
			Carrier number& $32$ & $8$ \\
			UE Speed & (10,20,40)m/s & (30,60,120)km/h \\
			Sampling interval & (1,2,3,4,5)ms & (1,2,3,4,5)ms \\
			Training Samples & $50000$ & $50000$ \\
			Testing Samples & $5000$ & $5000$ \\
			\toprule
		\end{tabular}
	\end{center}
	\label{experiment_setting}
\end{table}

Considering that the channel sampling time intervals are typically on the order of milliseconds in a practical system, in the ray-training scenario, we assume that the UE performs uniform linear motion in a random direction within the sequence sampling period. Then, the static CSI matrix of each sampling point is generated using the 3D ray-tracing algorithm. Further, the corresponding Doppler shift is calculated and applied to each propagation path, and finally, we obtain a series of the mobile channel sequences as the training and validation datasets.

\subsubsection{Benchmarks and Training Settings}
To comprehensively evaluate our proposed approach, we select three representative and competitive channel prediction methods as comparative benchmarks, including LSTM, Transformer proposed in \cite{9832933}, and ODE-RNN proposed in \cite{9978073,9857844}. To ensure fair comparisons, we use the same training settings for all schemes and keep the scale of the neural network's parameters roughly equivalent. NMSE (normalized mean square error) between prediction and ground truth is set as the performance indices.

\subsection{Experimental Results} \label{performance}
\begin{figure}[htb!]
		\centering
		\includegraphics[width=0.25\textwidth]{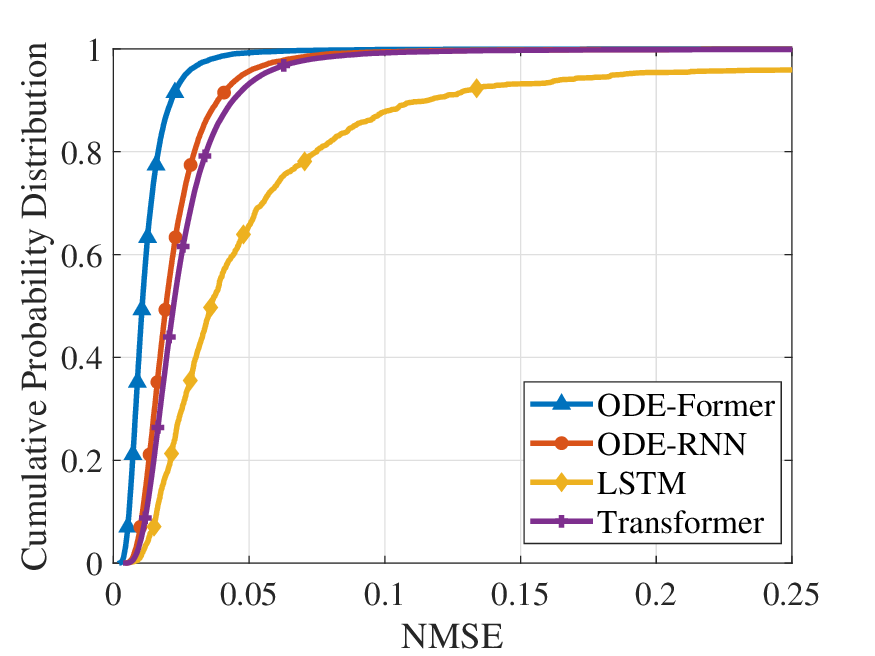}
		\caption{Cumulative probability distribution of errors between the acquired channel and the true channel (NMSE).}
		\label{exp_cdf}
		\vspace{-1em}
	\end{figure}

In order to intuitively compare the prediction accuracy of the proposed learning network and benchmarks, Fig. \ref{exp_cdf} shows the cumulative probability distribution of errors of the ray-tracing scenario. For the fairness of comparison, we choose the typical setting where the length of the sampled CSI sequence is $5$ and the CSI of the next time slot is to be predicted. The sampling interval is $1$ms, and UE's velocity is $10$m/s. It can be seen from the results that different from methods, such as LSTM and Transformer, that discretely generate the predicted CSI, the ODE-RNN can achieve better prediction accuracy. However, due to the inherent weakness of the network structure, the performance improvement compared to the Transformer is not significant enough. As a comparison, our proposed ODE-Former extensively makes up the existing weakness and shows excellent performance gain over other benchmarks.

\begin{figure}[htb!]
	\centering
	\subfigure[Ray-tracing scenario]{
		\begin{minipage}[b]{0.24\textwidth}
			\centering
			\includegraphics[width=1\textwidth]{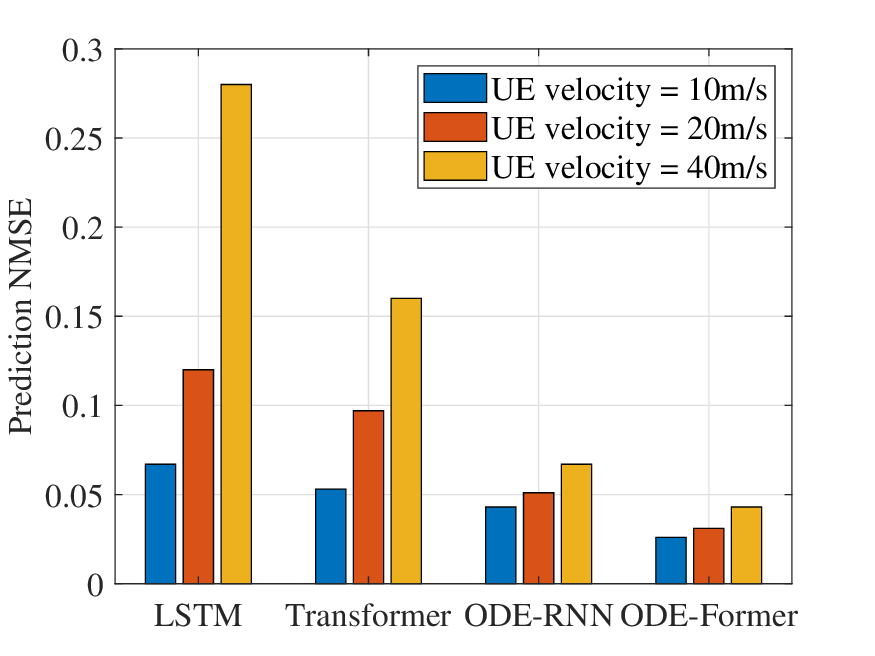}
			\label{different_velocity_simulation}
		\end{minipage}
	}
	\hspace{-0.6cm}
	\subfigure[Real-world scenario]{
		\begin{minipage}[b]{0.24\textwidth}
			\centering
			\includegraphics[width=1\textwidth]{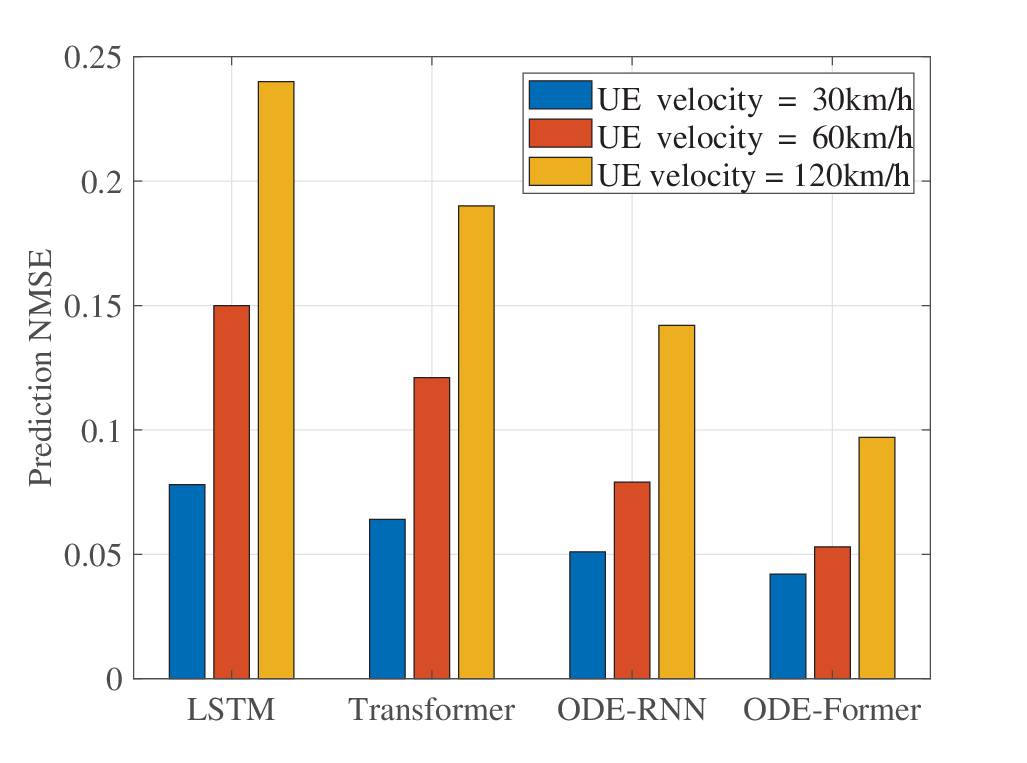}
			\label{different_velocity_real}
		\end{minipage}
	}
	\caption{The prediction NMSE comparison among different methods with various UE velocities.}
	\vspace{-1.5em}
\end{figure}

\begin{figure}[htb!]
	\centering
	\subfigure[Ray-tracing scenario]{
		\begin{minipage}[b]{0.24\textwidth}
			\centering
			\includegraphics[width=1\textwidth]{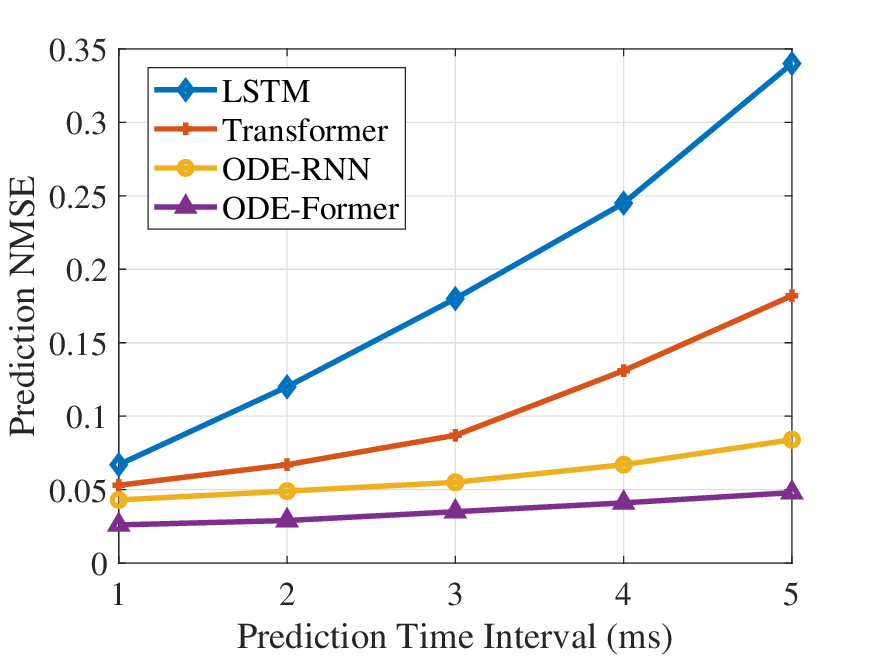}
			\label{different_interval_simulation}
		\end{minipage}
	}
	\hspace{-0.6cm}
	\subfigure[Real-world scenario]{
		\begin{minipage}[b]{0.23\textwidth}
			\centering
			\includegraphics[width=1\textwidth]{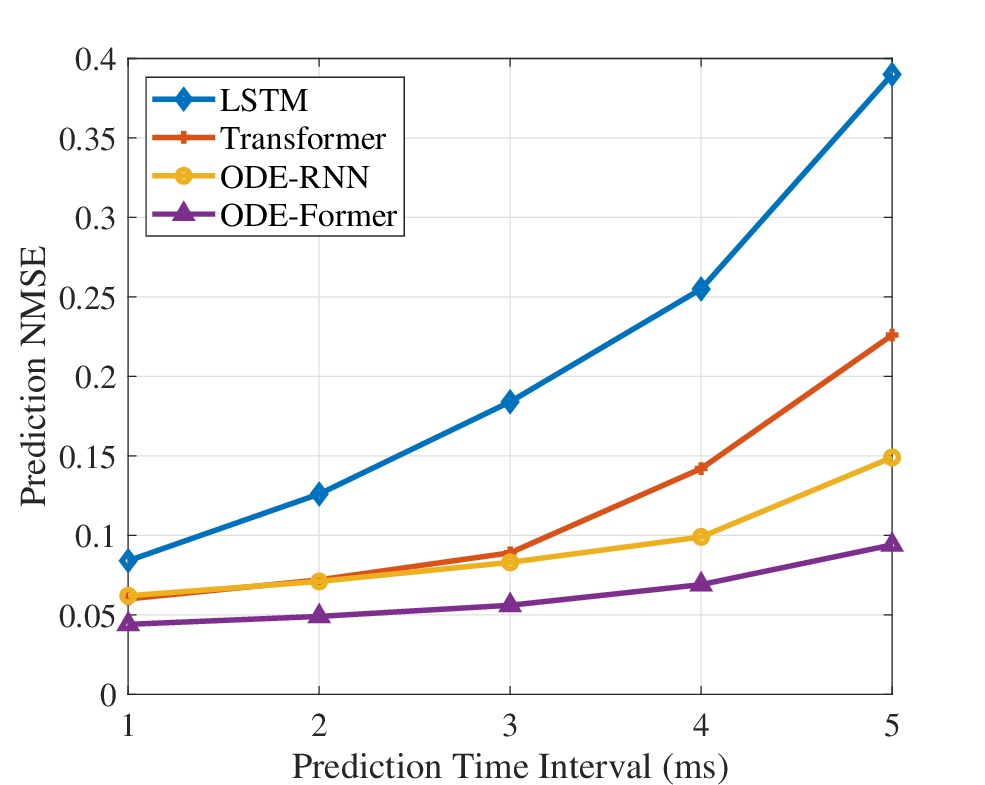}
			\label{different_interval_real}
		\end{minipage}
	}
	\caption{The prediction NMSE comparison among different methods with various prediction times intervals.}
		\vspace{-1.5em}
\end{figure}

\begin{table}[h!]
	\caption{complexity and inferencing time}
	\scriptsize
	\begin{center}
		\begin{tabular}{ p{2.5cm} p{0.8cm} p{1.2cm} p{0.8cm} p{0.8cm} }
			\toprule
			\textbf{Parameters} & \textbf{ODE-Former} & \textbf{Transformer} & \textbf{LSTM} & \textbf{ODE-RNN}\\
			\toprule
			Computation Complexity (MFLOPs) & $163.21$ & $63.12$ & $62.97$ & $103.12$\\
			Model Complexity (M) & $6.14$ & $6.29$ & $6.27$ & $6.18$\\
			Inference Time (ms)& $0.52$ & $0.23$ & $0.95$ & $4.13$\\
			\toprule
		\end{tabular}
	\end{center}
	\label{network_settings}
\end{table}

The robustness and feasibility of handling high-mobility scenarios constitute a pivotal aspect in evaluating channel prediction methodologies. To demonstrate the superiority of our proposed method, we compared its performance with competing approaches under various settings of UE velocities. As shown in Fig. \ref{different_velocity_simulation} and Fig. \ref{different_velocity_real}, it can be seen that the performance of existing works, such as LSTM and Transformer, drops significantly when switches to the high mobile scene. The main reason behind this is that those networks merely exploit limited correlations through discrete sampling points. However, as the UE's moving speed increases, the spatial distance between two sampling points under the same temporal interval becomes larger, resulting in a significant decrease in correlation. In contrast, our proposed scheme significantly outperforms existing methods across all velocity settings. Moreover, it maintains excellent performance even in high mobile scenarios, ensuring applicability in practical wireless systems.

Further, in order to better illustrate the advantages of the proposed network, Fig. \ref{different_interval_simulation} and Fig. \ref{different_interval_real} compare the prediction NMSE with various time intervals between adjacent sampling points of two scenarios. It can be seen from the experiment results that due to the weakening of correlation caused by longer time intervals, the prediction accuracy of the transformer and LSTM decreased significantly. In comparison, the proposed ODE-Former network can deal with longer time intervals without losing much performance.

Moreover, to prove the proposed method's scalability and feasibility, we compared several critical parameters associated with model inference. It is noteworthy that all models run on Nvidia Tesla A800 GPU. In PyTorch, the Transformer model benefits from automatic parallel computation. For our model, we specifically develop parallel acceleration and pipeline codes, which include the parallel processing of implicit components extraction and assemble blocks, multi-sampling point, and multi-head parallelism within the multi-samples fusion block, as well as pipeline parallelism for the stacked ODE-Former blocks. As shown in Table \ref{network_settings}, under the condition of similar overall model complexity, the ODE-RNN and LSTM networks of the comparative solution, despite having lower computational complexity, suffer from low utilization of modern computing hardware due to the sequential network structure, resulting in significant inference latency. However, the proposed ODE-Former network, although it needs more computation resources, achieves inference latency close to that of the Transformer network because most of its components can be computed in parallel, which ensures the feasibility of the proposed approach.

\section{Conclusions} \label{conclusion}
This letter proposed an innovative ODE-Former network for mobile channel prediction. Through unique network design, this proposed method significantly extended the learning capability and the ability to characterize physics process-driven functions of the original Neural ODE with a former-like design. With the proposed network, the system could break free from uniform sampling and serial prediction constraints. Experiments of comparison of state-of-the-arts verified that the proposed method could significantly enhance prediction accuracy, system flexibility, and network robustness compared with existing methods.

\bibliographystyle{IEEEtran}
\bibliography{bibfile}
\end{document}